\documentclass[apl,twocolumn,a4paper]{revtex4}
\usepackage{graphicx}

\begin{document}

\title{Suspended single-electron transistors: fabrication and measurement}

\author{G.~S.~Paraoanu and A.~M.~Halvari}
\affiliation{NanoScience Center, Department of Physics, University of Jyv\"askyl\"a,
P.O.~Box 35 (YFL), FIN-40014 University of Jyv\"askyl\"a, FINLAND}

\begin{abstract}

We have fabricated aluminum single-electron transistors in which the island is not in contact
with the substrate. This new type of device, which can be called suspended single-electron transistor
(SUSET), displayed well-defined I-V and dI/dV-V features typical for high-quality standard SET's.

\end{abstract}

\maketitle


The single-electron transistor (SET), a device proposed by Averin and Likharev in
1986 and
first built by Fulton and Dolan in the next year \cite{averinlikharev}, consists of a small
metalic island connected to two leads (electrodes) by two tunnel junctions. The device
is controlled through the voltage applied on a gate electrode capacitively coupled to the
island.

The SET was proposed for a variety of applications,
including counting electrons directly for a capacitance standard \cite{electrometer},
direct measurements of
single electron tunneling and co-tunneling, Bloch oscillations \cite{blochoscillations},
and microwave photon detection \cite{sawphoton}. There has been a lot of interest
recently in the use
of this device or relatively similar ones as quantum bits for a solid-state quantum computer \cite{squbits} or
as read-out systems for quantum circuits \cite{readouts}.
Finally, multiple-island devices such as Cooper pair pumps and single-electron turnstiles
could find applications as metrological standards for current \cite{currentstandard}
or for measuring decoherence
times in nanoscale superconducting systems \cite{jukka}.

The Hamiltonian of an SET contains typically two parts: a tunneling Hamiltonian and a
charge Hamiltonian.
The first one can be either single-electron tunneling, if the SET is operated in the
normal regime, or Josephson tunneling, in the case of a superconducting SET
(Cooper-pair transistor). The charge Hamiltonian is given
by the total energy associated
with the electrostatic energy of the island plus the work done by sources to move
the charge around the circuit.
This energy contains also a phenomenological quantity usually called
"offset charge", which brings an extra contribution to the energy even at zero
voltages applied. This quantity takes different values for
each device and cooldown and  it fluctuates in time with an 1/f spectral density.

This effect limitates dramatically the usefulness of the device in many applications,
for example it is the main constraining factor for the sensitivity of low-frequency SET
electrometers (currently of the order of
$\times 10^{-5}$e/$\sqrt{\mathrm{Hz}}$ \cite{lowfelectrometry}); the r.f.-SET
improves this figure by paying the price of going up to higher frequencies, thus
making the device more difficult to operate.

It is believed \cite{quantumamplifier} that this phenomenon is
caused by the trapped charge fluctuators
in the oxide junction and the substrate;
further careful investigations have led to the conclusion that the noise coming from the
substrate dominates \cite{1/f}.

One idea to overcome these problems would be to try diferent substrates than the usual SiO$_2$; indeed,
the existing noise measurements in SET's seem to suggest that there is a correlation between
the type of substrate used and the minimum sensitivity achieved.
This paper proposes a more radical approach to this issue:
to eliminate the
substrate itself. We demonstrate that it is indeed possible to fabricate working
high-quality devices Al superconducting SET's in which the
island is not in contact with the substrate.

Although related free-standing structures have been fabricated before,
most notably with
semiconducting and metallic nanowires and carbon nanotubes \cite{suspended},
active devices such as superconducting transitors have been available only on membranes or other types of specially
designed supports. Besides immediate applications for present mesoscopic electrical circuits
requiring low levels of substrate noise, one
can speculate that these devices are relevant for building future hybrid solid state - atom (ion)
systems which were recently proposed in the context of quantum computing \cite{zoller}.

\begin{figure}[htb]
\includegraphics[width=84truemm]{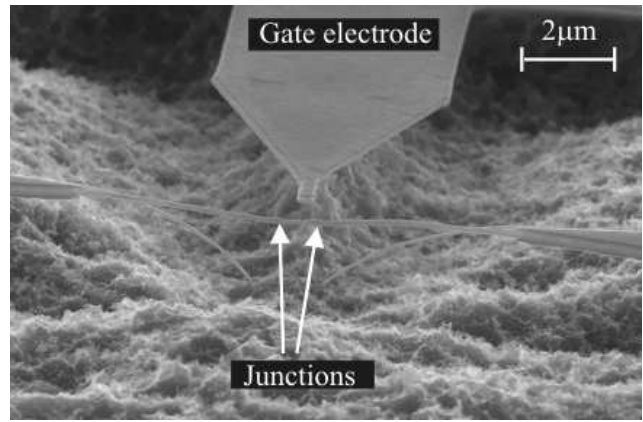}
\caption{High-magnification SEM image of an Al SUSET. The tip of the gate electrode (center up)
points to the island. The junctions are barely visible in this picture as two thicker nodular
structures along the suspended aluminum wire. The two lines starting from the left and
right electrodes and pointing downwards (toward the substrate) are the usual
by-product of two-angle evaporation. }

\label{sharpest}
\end{figure}


The suspended Al SET was fabricated using conventional two-angle evaporation
on a nitridized Si wafer of 300 nm thickness followed by reactive ion etching.
A double layer resist of PMMA-P(MMA-MAA) was spinned on
the top of the SiN$_2$ substrate and baked at 160${^\circ}$C for 45
min (the bottom layer) and 60 min (the upper one).
The resulting thicknesses of the PMMA and P(MMA-MAA)
were approximately 250 nm and 350 nm, respectively.
The mask was drawn using a scanning electron microscope (JEOL, JSM 840A)
with acceleration voltage of 20 kV. The minimum width of the pattern in the island region was 50 nm;
after the whole process, the thickness of the metalic wires increased to 100 nm.
Then the resists were developed in a mixed solution (1:2)
of methyl-{\em iso}-butylketon (MIBK)
alcohol with isopropanol for about 30 s (the upper layer of PMMA) followed immediately by rinsing in isopropylic alcohol,
and then in a mixture
(1:2) of methyl glycol and methanol for roughly 8 s (the lower layer of P(MMA-MAA)).
The chip was further cleaned in a reactive ion etcher
(AXIC BENCHMARK) at 30 mTorr pressure with 50 sccm flow of O$_2$
and 48 W RF-power for 30 s, a process that removes the resist leftovers from the SiN$_2$ surfaces.
The aluminum was evaporated in an UHV chamber at pressure of  2--4$\times 10^{-8}$ mbar
with an evaporation rate of 0.5 nm/s for both the island and the electrodes.
The film, with a thickness of 40 nm, was then oxidized
in a steady flow of O$_2$, at a pressure of 20 mbar for three minutes. The junctions
were formed by depositing another layer of 60 nm Al at a different angle. In the sample presented here,
the 40 nm and 60 nm correspond to the island and the leads, respectively.
The process was completed by lift-off, with the sample being slowly heated in acetone
up to the boiling point (56$^{^\circ}$C).

Finally the SET was put into a RIE (AXIC BENCHMARK) and etched in a two step process. The power
was 10\% (60 Watts), and the flow of O$_2$ and CF$_4$ was 10 \% (5 sccm) and 50\% (10.5 sccm)
respectively. The first step was anisotropic etch at a pressure of 40 mTorr for 3 minutes (ion
bombardment to break the surface of the nitride). The second step (isotropic etch)
was at a pressure of 70 mTorr for 7 minutes.

\begin{figure}[htb]
\includegraphics[width=84truemm]{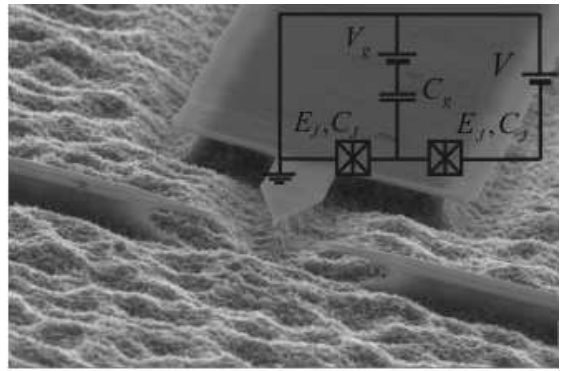}
\caption{A lower-magnification SEM picture of a suspended single-electron transistor, showing
the effect of etching on the SiN$_2$ substrate. The structure doesn't touch the resist for
a length of about 20 $\mu$m. The inset in the upper right corner shows the full large-scale
structure, including the bonding pad for the gate.}
\label{larger}
\end{figure}

The resulting structure is shown in Fig. \ref{sharpest} and Fig. \ref{larger}.
The length of the island is about 1$\mu$m, and the thickness is approximately 100 nm.

\begin{figure}[htb]
\includegraphics[width=84truemm]{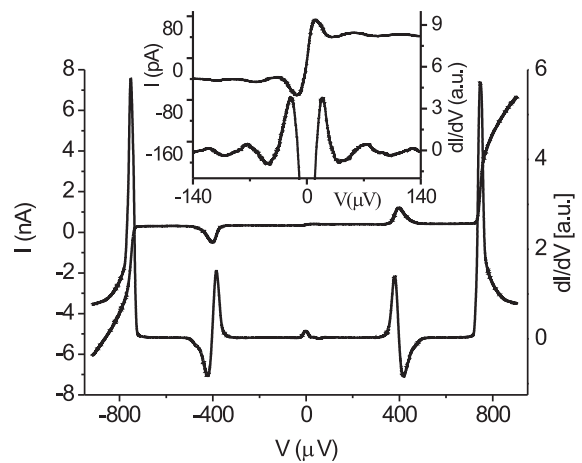}
\caption{Current and conductance as function of bias voltage, showing clearly the defining features
of a good-quality SET (Josephson current, JQP, and quasiparticle conductance peak). The inset is a
low-bias voltage detail, with a better visible Josephson effect and mild Cooper pair resonant peaks
in the current.}
\label{current_cond}
\end{figure}

The SUSET was measured in an electrically shielded room using a Nanoway PDR-50 dilution refrigerator
with the lines filtered with commercial $\pi$-filters at 4.2 K and a combination of $\pi$-filters and
RC-filters at room temperature. The conductance was measured using a lock-in amplifier technique.
The results are presented in Fig. \ref{current_cond}. The sample presented here had a room-temperature
resistance $R_n = 73$k$\Omega$.

\begin{figure}[htb]
\includegraphics[width=84truemm]{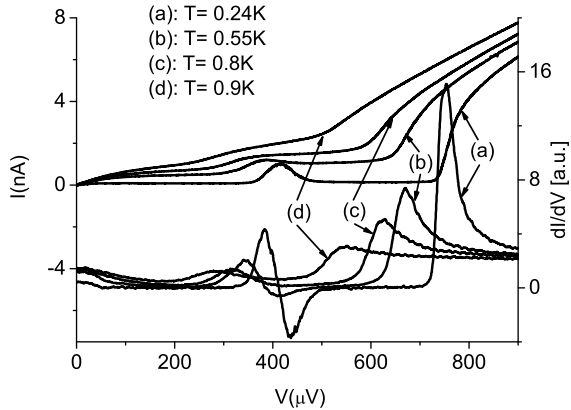}

\caption{I-V characteristics as a function of temperature. The sharpest features correspond to
T = 0.2 mK; the temperature was then raised close to the critical temperature of Al.
\label{Temp}}
\end{figure}

From the threshold voltage of the quasiparticle branch at 0.755 meV we can determine the
gap of Al ($\Delta_\mathrm{Al} = 0.189$ meV). The charging energy of the sample was derived from the Coulomb-blockade
conductance measured at 4.2 K; we found a change in the zero-bias conductance peak
$\delta G /G_{T} = 1.24\%$,
yielding a charging energy $E_{c}=e^{2}/2C_{\Sigma}= 13.5\mu$eV, where
$C_{\Sigma} = C_{1} + C_{2} + C_{g}$ is the
usual definition for the total capacitance of the island.

The critical Josephson current from Fig. \ref{current_cond} is approximately 70 pA;
switching-current measurements, which are more precise, gave a value  $I_c = 61$ pA for this sample.
This corresponds to a quite small
Josephson energy $E_{J}= \hbar I_{c}/2e = 0.13 \mu$ eV. The Ambegaokar-Baratoff formula
$E_{J} = R_{K}\Delta_\mathrm{Al}/8 R_{n}$, where $R_K = h/e^2 = 25.8$k$\Omega$ is the quanta of resistance
(von Klitzing constant), yields $8.3 \mu$ eV, significantly larger than the measured value.
These are a well-known discrepancies for small-junction superconducting SET's,
caused by the sensitivity of the Josephson effect
for small junctions to the
external electromagnetic environment \cite{JJ}. They occur when $E_{c}$ is
larger or of the same order of magnitude as $E_{J}$ (therefore the phase of the island
is not a good quantum number and the current has strong fluctuations), which is indeed
the case for our sample.

In Fig. \ref{current_cond} the two symmetric peaks in the current are due to a combined
Josephson and quasiparticle (JQP) tunneling in the junctions; this effect is predicted to
happen at a bias voltage
$2 \Delta_\mathrm{Al} + E_\mathrm{C} \leq eV \leq
2 \Delta_\mathrm{Al} + 3E_\mathrm{C}$ \cite{jqp}. For our sample, this gives
0.391 meV$\alt eV \alt$ 0.418 meV, in excellent agreement with the position of the peak in the
experimental data. The JQP features are quite broad, as expected
from the theory of this process for samples with small $E_c /\Delta$ \cite{JQP-Tinkham}.

Finally, a measurement at low bias voltages (inset of Fig. \ref{current_cond}) reveals a
finer structure of resonances in the current, corresponding to oscillations in the conductance.
According to the theory of resonant tunneling of Cooper pairs \cite{resonances}, the spacing between consecutive peaks
peaks should be $4E_\mathrm{C}$, i.e. 54 $\mu$eV. Although the data were quite noisy, it was still
possible to estimate the distance between peaks as being indeed 50-60 $\mu$eV.

The SUSET is affected by temperature in the same way as conventional SET's.
Fig.~\ref{Temp} presents I-V measurements from 0.2 K to about 0.9 K, close
to the critical temperature of Al. The low-temperature features get more rounded and they are
displaced towards lower bias voltages, in agreement with the decrease of the gap of Al.


In conclusion, we have fabricated and measured a suspended SET. We did not observe any decrease
in the quality of the device due to the extra etching process. Therefore, we believe that this
new device can be useful in applications that require a precise control of the
island charging as well as for investigations of the origin of the noise in
single-electron transistors.


G.~S.~P. was supported by an EU  Marie Curie Fellowship (HPMF-CT-2002-01893); A.~P.~H. was
supported through the SQUBIT project (IST-1999-10673) of the European Union and the Academy of Finland
Center of Excellence in Condensed Matter and Nuclear Physics at the University of Jyv\"askyl\"a.

\end{document}